\documentstyle[11pt]{article}
\begin{document}
\title{
Dynamics   of   metastable   vortex   states   in  weakly  pinned
superconductors: A phenomenological model\\
\vskip 0.25truecm
\normalsize
\noindent
G. Ravikumar
\vskip 0.1truecm
{\it
Technical Physics and Prototype Engineering Division,
Bhabha Atomic Research Centre,
Mumbai-400085, India\\}
\vskip 0.2 truecm
}
\maketitle
\normalsize
\normalsize

\begin{abstract}
We  present  a  phenomenological model for the vortex dynamics in
the  peak  effect  region  of  weakly  pinned superconductors. We
explain the history dependent dynamic response of the  metastable
vortex states subjected to a transport current and the hysteretic
voltage-current  characteristics observed in the vicinity of peak
effect in weakly pinned superconductors. A  strong  variation  in
voltage  current  characteristics with the current sweep rate and
the anomalous dependence of critical current density $J_c$ on the
magnetic field sweep rate have also been accounted  for  by  this
model.

\end{abstract}
\noindent
In   weakly   pinned  superconductors,  the  competition  between
intervortex interaction and quenched disorder  produces  a  sharp
peak  in the critical current density $J_c$ just below the normal
state   boundary.   This   phenomenon,   known   as   the    peak
effect\cite{higgins} occurs when the vortex lattice passes from a
state  with a high degree of spatial order to an amorphous pinned
state \cite{larkin,gammel}. The peak effect is often  accompanied
by   a  marked  history  dependence  in  $J_c$  originating  from
different metastable configurations in which the  vortex  lattice
could                                                       exist
\cite{steingart,wordenweber,henderson,fcrev,supercool}.

Another  aspect  of  the  metastability  associated with the peak
effect is seen in the history dependent dynamic response  of  the
vortex  lattice  to  a  transport  current\cite{henderson}. It is
presumed that this behavior is due to the  rearrangement  of  the
vortex lattice from one metastable configuration to another aided
by an external driving force\cite{xiao}. Moreover, the process of
rearrangement  can  be  halted  or  in  other words, a particular
metastable  configuration  can  be  frozen  in  time,  by  merely
switching  {\it  off} the transport current\cite{xiao}. Switching
{\it on} the current again after  a  time  interval  revives  the
further  evolution  of  the  metastable vortex configuration from
where it was halted (long term  "memory"  effect).  Further,  the
evolution   of   a   metastable   vortex   configuration  can  be
considerably slowed  down  by  slightly  lowering  the  transport
current\cite{xiao}.

Additionally,  Zhukov  {\it  et  al}\cite{zhukov}  have  recently
reported  a  magnetic  field  sweep  rate   ($dB/dt$)   dependent
magnetization  study in the peak region of $2H-NbSe_2$ and in the
{\it fishtail} region of $YBa_2Cu_3O_7$. Their results imply that
in the increasing field cycle $dJ_c/d|dB/dt|  <  0$  at  a  fixed
applied field. This so called "negative dynamic creep" phenomenon
is  in  contast  to  the  behavior  expected  from  the thermally
activated  flux  creep  where  $J_c$  is  expected  to   increase
logarithmically   with   the   sweep   rate   ($dJ_c/d|dB/dt|   >
0$)\cite{pust}. However, in the decreasing  field  cycle,  it  is
found that $dJ_c/d|dB/dt| > 0$.

In  transport  experiments,  $J_c$  is  identified as the current
density  $J$  at which the electric field $E$ exceeds a threshold
value. Usually pinning  properties  of  the  vortex  lattice  are
uniquely  characterized  by the $J_c$ value and the corresponding
$E-J$ relation. However, in the peak region of $2H-NbSe_2$, $J_c$
is found to increase significantly with the rate at which $J$  is
increased\cite{xiao}.  This  amounts  to  the  $J_c$ vs $T$ curve
being  much  sharper  when  a   large   value   of   $dJ/dt$   is
used\cite{xiao}.   Moreover,   $E-J$  curves  themselves  exhibit
significant   hysteresis   in   upward   and   downward   current
cycles\cite{steingart,wordenweber,henderson}.

The  history  dependence  of  $J_c$  under different quasi-static
field  excursions  was  recently  explained by a phenomenological
model\cite{model}. This model supposes that there exists a set of
metastable   vortex   configurations,  each  corresponding  to  a
different value of $J_c$.  At  a  given  field  and  temperature,
different  configurations  differ  only  in the extent of lattice
order. $J_c$ can thus be viewed as a  macroscopic  representation
of   a  particular  metastable  vortex  configuration.  The  most
important assumption of this model is that there  exists  a  {\it
stable}  vortex  configuration  with  $J_c  =  J_c^{st}$ which is
unique for a  given  field  and  temperature.  Depending  on  the
particular  magnetic  history,  a  metastable  configuration with
$J_c$ either smaller or larger than $J_c^{st}$ is  produced.  The
extent  of  history  dependence  is assumed to be governed by the
parameter  $B_r$,  which  is  a  macroscopic   measure   of   the
metastability. Evolution of $J_c$ when the field changes from $B$
to $B + \delta B$ is postulated to be governed by

$$
J_c(B+\delta B)  =  J_c(B)  +  (|\delta  B|/B_r)  [J_c^{st}(B)  -
J_c(B)]
\eqno{(1)}
$$

\noindent
Eq.  1  provides  that  a  metastable configuration with $J_c \ne
J_c^{st}$ would evolve into the stable state when  the  field  is
repeatedly  cycled  by a small amplitude. Recently, the existence
of    such    a    stable    state     has     been     confirmed
experimentally\cite{cycle}.  Physically,  we can imagine that the
process of repeated field cycling pumps the vortices in  and  out
of  the superconductor. In the absence of adequate thermal energy
field  cycling  allows  the  vortex  lattice   to   explore   the
configuration  space and reorganize or {\it anneal} into the most
stable vortex configuration. Thus the process of {\it  annealing}
is  aided  by  the  motion of vortices caused only by an external
driving force such as a  field  change  or  a  transport  current
larger than $J_c$\cite{xiao,paltiel}.

We  believe  that  the dynamic phenomenon mentioned above and the
history  dependence  in  critical  currents  are  two   different
manifestations  of  the  same  phenomenon of metastability in the
vortex  state  and  therefore  should  be  described  by a common
phenomenology. In this paper, we present a generalization of  the
phenomenological  model  described  above, to account for (i) the
history dependent dynamic response of the vortex lattice  in  the
peak  region  (c.f  Fig.  4  of  Ref. \cite{henderson}); (ii) the
hysteresis              in              $E-J$              curves
\cite{steingart,wordenweber,henderson};  (iii)  the dependence of
$J_c$   on   the   current   sweep    rate    used    in    $E-J$
measurements\cite{xiao}  and  (iv)  the  so  called {\it negative
dynamic creep} phenomenon\cite{zhukov}.

We  postulate  that,  the  time  evolution of a metastable vortex
configuration in terms of $J_c$ is described by the equation

$$
dJ_c/dt = (J_c^{st} - J_c)/\tau
\eqno{(2)}
$$

\noindent
where  $t$  is  the time variable and $\tau$ is the time constant
for  the {\it annealing} process. We argue that the time constant
$\tau$ is inversely related to the velocity $v$ of vortex motion,
i.e.,        $\tau$        $\sim$        $|v|^{-1}$        $\sim$
$|E|^{-1}$\cite{paltiel,shobo}. In other words, a larger velocity
of vortices facilitates faster rearrangement of vortices into the
{\it stable} configuration (annealing). The absolute value of $E$
(or $v$) implies that the process of rearrangement is independent
of the direction of vortex motion.

Let  us  first  examine  the  vortex  rearrangement  driven  by a
quasi-static field sweep, in the absence of transport current. If
the field sweep rate is $dB/dt$, $E  \sim  dB/dt$  and  the  time
constant  $\tau$  ($\sim  |E|^{-1}$)  can be expressed as $\tau_0
R/|dB/dt|$  where $R$ is a characteristic value of the sweep rate
below which the field change can be considered quasi-static.  The
time  constant  $\tau_0(B,T)$  can be a function of the field $B$
and temperature $T$. Substituting the above expression for $\tau$
in Eq. 2, we get

$$
dJ_c/dB = \pm (J^{st}_c - J_c)/B_r
\eqno{(3)}
$$

\noindent
where  $B_r  =  \tau_0  R$  with upper (lower) sign for the field
increasing (decreasing) case. Eq. 3 is the differential  form  of
Eq.    1    describing    the    magnetic    history    dependent
$J_c$\cite{model}. Therefore,  we  conclude  that  the  parameter
$B_r$  which  is a measure of the history dependence in $J_c$, is
directly related to the time constant $\tau_0$ which, as would be
shown below, governs the dynamics  of  metastable  vortex  states
subjected to an external driving force.

We  now  consider the dependence of $J_c$ on magnetic field sweep
rate  for  a  fixed applied field. When the external field $B(t)$
increases (or decreases) linearly in time, we can write $t =  \pm
[B(t) - B_i]/\dot{B}$. The upper (lower) sign is again applicable
for  increasing  (decreasing)  field. $B_i$ is some initial field
where the field sweep started and  $\dot{B}$  (positive)  is  the
magnitude  of the sweep rate $dB/dt$. Substituting for $t$ in Eq.
2, we get

$$
dJ_c/d\dot{B} = \mp [(B(t) - B_i)/\dot{B}^2](J^{st}_c - J_c)/\tau
\eqno{(4)}.
$$

\noindent
In  the  increasing  field  cycle  (upper  sign)  $(B(t)-B_i)$ is
positive    and    $J_c    <    J_c^{st}$     in     the     peak
region\cite{model,cycle}. Therefore $dJ_c/d\dot{B} < 0$ (negative
dynamic  creep)  as  observed in experiments\cite{zhukov}. On the
other  hand,  in  the  decreasing  field   cycle   (lower   sign)
$dJ_c/d\dot{B}   >   0$   because   $J_c   >  J_c^{st}$(see  Ref.
\cite{model}) and $(B(t)-B_i)$ is negative. In other  words,  the
{\it  negative  dynamic  creep}  observed in the increasing field
cycle\cite{zhukov} is a manifestation of the  finite  time  delay
needed for the vortex lattice to become more disordered.

We  now  focus  on  the  transport current driven {\it annealing}
process governed by Eq. 2 where  the  time  constant  $\tau  \sim
|E(J)|^{-1}$.  For  $E(J)$,  we choose the simple form $E(J) \sim
(J/J_c - 1)$ for $J > J_c$ and thus

$$
\tau(J) = \tau_0/(J/J_c - 1)
\eqno{(5)}.
$$

\noindent
For  $J  <  J_c$,  $E(J)  =  0$  and  therefore $\tau \rightarrow
\infty$. The specific form  chosen  for  $E(J)$  is  however  not
critical  for the results we are going to present. Any other form
which makes $\tau \rightarrow  \infty$  as  $J  \rightarrow  J_c$
would equally serve to illustrate the results in this paper.

Eq.  2  and Eq. 5 suggest that irrespective of whether (i) $J_c <
J_c^{st}$  or  (ii)  $J_c  >  J_c^{st}$, a transport current $J >
J_c$  causes  $J_c$  to increase or decrease respectively towards
the stable value. In other words, while  $J  >  J_c$,  $J_c$  and
thereby $\tau$ change with time, resulting in the time dependence
of the voltage response through the relation $E(J) \sim (J/J_c(t)
- 1)$, even when the transport current $J$ is kept fixed.

Combining  Eq.  2  and  Eq.  5 and substituting the dimensionless
quantities $j_c(t) = J_c(t)/J_c^{st}$, $j =  J/J_c^{st}$  and  $t
\rightarrow  t/\tau_0$,  a  formal  solution  for $j_c(t)$ can be
written as

$$
j_c(t) = 1 + [j_c(t_0) - 1]\exp[-\int_{t_0}^{t} dt (j/j_c(t) - 1)]
\eqno{(6)}
$$

\noindent
where $j_c(t_0)$ is the normalized critical current of the vortex
configuration  in  which the system is initially prepared at time
$t_0$. Here we have two cases of interest viz., (i)  $j_c(t_0)  <
1$  and  (ii)  $j_c(t_0) > 1$ which are henceforth referred to as
{\it superheated} and  {\it  supercooled}  vortex  configurations
respectively\cite{cycle}. The {\it superheated} configuration can
be  achieved by entering the peak region either by increasing the
field at constant temperature\cite{model,cycle} or by  increasing
temperature  at  a  constant  field.  On the other hand, the {\it
supercooled} state is achieved by  either  decreasing  the  field
from   above   the   normal  state  boundary  while  keeping  the
temperature constant\cite{supercool,model,cycle}  or  by  cooling
the superconductor in a field\cite{fcrev,supercool}.

We  solve  Eq.  6  iteratively  to  obtain  the time evolution of
$j_c(t)$ in response to a step increase in the transport  current
density  $j$  at  $t  =  t_0$.  In  Fig.  1,  we present the time
dependence of $j_c(t)$ and $E(t) \sim (j/j_c(t)  -  1)$  for  the
{\it  superheated} vortex configuration, i.e, $j_c(t_0) < 1$. The
voltage response decreases with time as  observed  experimentally
(c.f.  Fig.  4(b)  of  Ref.  \cite{henderson}  and Fig. 2 of Ref.
\cite{xiao}). In the inset  of  Fig.  1(b),  we  show  $E(t)$  vs
$ln(t)$  at  different  driving  currents,  which is in excellent
qualitative  agreement  with  the experimental results (c.f. Ref.
\cite{xiao}). In Fig. 2, we present $j_c(t)$ and $E(t)$  obtained
by  starting from a {\it supercooled} vortex configuration (i.e.,
$j_c(t_0) > 1$), which is again  in  good  qualitative  agreement
with experiments(c. f. Fig. 4(b) of Ref. \cite{henderson}).

Let us now consider the situation in Fig. 1, but with the driving
current $j$ switched {\it off} to zero during the interval $t_1 <
t  <  t_2$.  At  time  $t_1$, the vortex lattice is frozen into a
metastable configuration corresponding to $j_c = j_c(t_1)$ as the
time constant $\tau \rightarrow \infty$ for $j < j_c$.  When  the
current  is switched {\it on} again at $t = t_2$, $j_c$ begins to
evolve from the value $j_c(t_1)$ according to Eq. 6 as  shown  in
Fig.  3(a).  Similarly, one can consider the case of reducing the
current $j$ by a small amount over a time interval  as  shown  in
Fig.  3(b).  This  merely  slows down the process of evolution of
$J_c$  towards  a stable value $J_c^{st}$. The calculated voltage
responses shown in Figs. 3(a) and 3(b) compare very well with the
results of Ref. \cite{xiao}.

The above discussion provides a framework for the {\it annealing}
process  mediated  by  a field change\cite{model,cycle}. When the
critical state is set up,  shielding  currents  (just  like  eddy
currents)    are    set    up   at   a   high   current   density
($>>J_c$)\cite{wilson} which decay to  the  value  $J_c$  with  a
characteristic  time  constant  $T \approx \mu _0 a^2/4\rho_{ff}$
governed by the flux flow resistance  $\rho_{ff}  =  \rho_n  H  /
H_{c2}$, where $\rho _n$ is the normal state resistivity, $H$ and
$H_{c2}$  are  the  external  field  and the upper critical field
respectively, $\mu_0 = 4\pi \times 10^{-7}$ W/A.m and $a$ is  the
sample  dimension.  This  is  equivalent  to a pulse of transport
current $J_{mean} > J_c$ of duration $T$ which causes  a  partial
annealing of the metastable vortex state.

This  model can also be used for calculating the $E-j$ relations.
We  consider  that the transport current density $j$ ($> j_c$) is
increased in small steps of $\Delta j$ each of  duration  $\Delta
t$. We calculate the $j_c$ and the corresponding $E(j)$ values at
the  end of each current step. In Fig. 4 we plot the $E-j$ curves
obtained with two different values  of  the  current  sweep  rate
$dj/dt$  ($\simeq  \Delta  j/\Delta  t$).  Clearly,  the critical
current  density inferred from these curves would be smaller when
$dj/dt$  is larger, just as observed in experimental studies(c.f.
Fig. 1 of Ref. \cite{xiao}). Further, the shape of  $E-j$  curves
differ dramatically with the current sweep rate.

We  also  studied  the  $E-j$  curves in both upward and downward
current cycles with a fixed $dj/dt$. In Fig. 5(a) we consider the
case  where  the  vortex  lattice  is  iniatially  prepared  in a
metastable configuration with $j_c(t_0) < 1$  ({\it  superheated}
state)  and  in  Fig.  4(b),  for  the  case $j_c(t_0) > 1$ ({\it
supercooled} state).  The  calculation  clearly  brings  out  the
hysteretic $E-J$ behavior usually seen in the peak region and the
results      compare      remarkably      well      with      the
experiments\cite{steingart,wordenweber,henderson}.   We   further
note that, after current cycling, $j_c$ ($J_c$) always settles to
the   value   1   ($J_c   =   J_c^{st}$)   as   shown   in   Ref.
\cite{wordenweber}.

The  results  presented  here are contrary to the notion that the
$E-J$  relation  and  the  $J_c$ value inferred from it, uniquely
characterize the pinning  properties  of  the  underlying  vortex
structure  in all circumstances. We argued instead that while the
transport current $J$ is higher than the threshold  value  $J_c$,
the  $J_c$  value  itself  changes  irreversibly,  leading to the
hysteresis in the $E-J$ relations.

We  caution  here that the simple model presented here explains a
variety of dynamic response of the vortex lattice to a  transport
current  {\it  above}  the  macroscopic  threshold  value  $J_c$.
However, it {\it can  not}  account  for  the  {\it  ac}  dynamic
response  for  current  values  below  $J_c$\cite{henderson2}.  A
qualitative mechanism for  the  {\it  ac}  dynamic  response  was
recently  presented by Paltiel et al\cite{paltiel} by considering
the disordered vortex phase, injected  into  the  superconducting
sample  across  the  surface  imperfections,  coexisting with the
ordered phase in the interior of the sample. However, an explicit
connection between this mechanism and the model developed here is
not yet available.

In  conclusion,  we  have  presented  a  macroscopic  model which
accounts for the dynamics of  the  vortex  lattice  in  the  peak
effect   region.  We  explained  the  history  dependent  dynamic
response of the metastable vortex states subjected to a transport
current  and  the  hysteretic   voltage-current   characteristics
observed in the peak region of weakly pinned superconductors such
as  $2H-NbSe_2$. We are also able to explain the strong variation
in voltage current characteristics with the current  sweep  rate.
Further,  the  so  called phenomenon of "negative dynamic creep",
viz., the anomalous dependence of $J_c$  on  the  magnetic  field
sweep rate in the peak region is also explained.

The author thanks Prof. S. Bhattacharya, Dr. K. V. Bhagwat, Prof.
A.  K.  Grover,  Prof.  S.  Ramakrishnan  and Dr. V. C. Sahni for
discussions and critical reading of the manuscript.

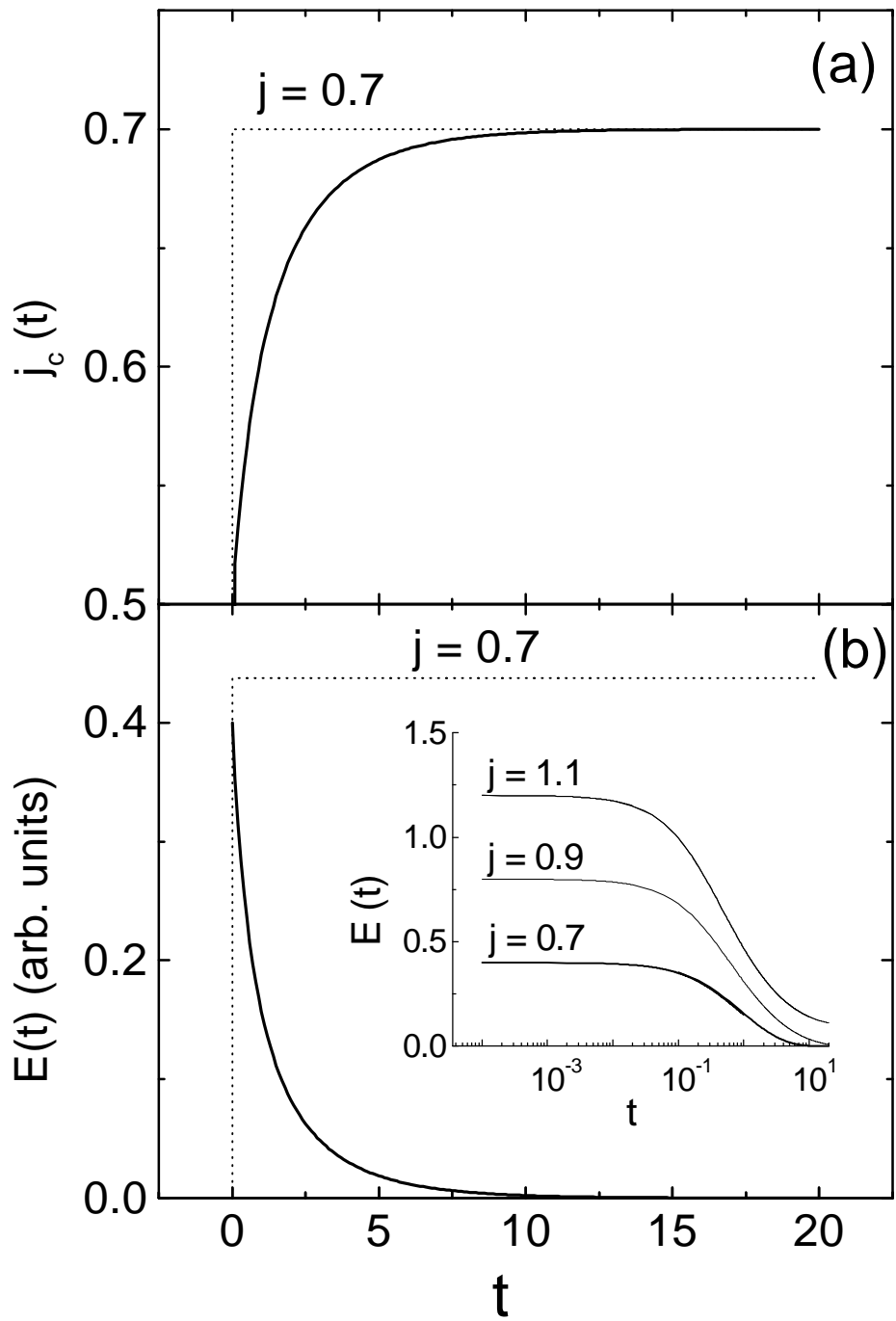
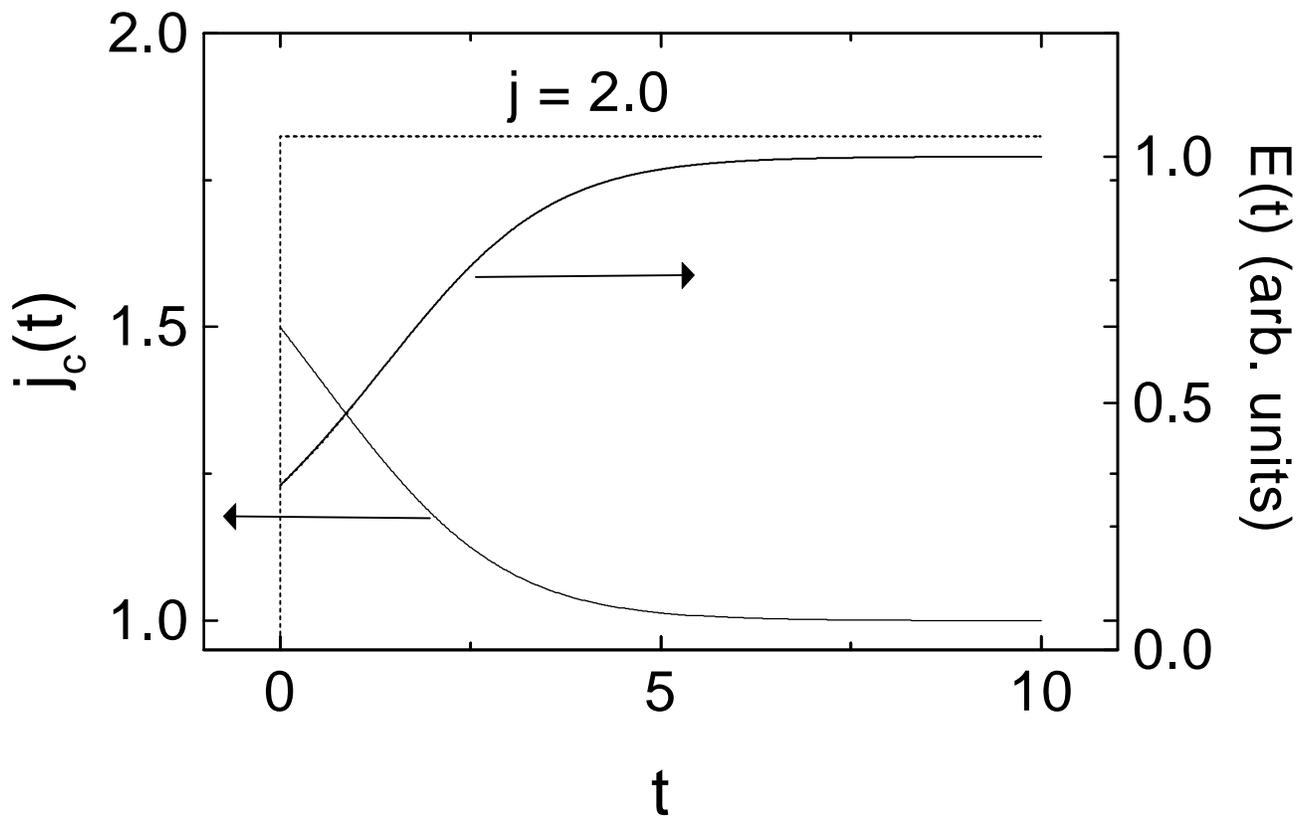
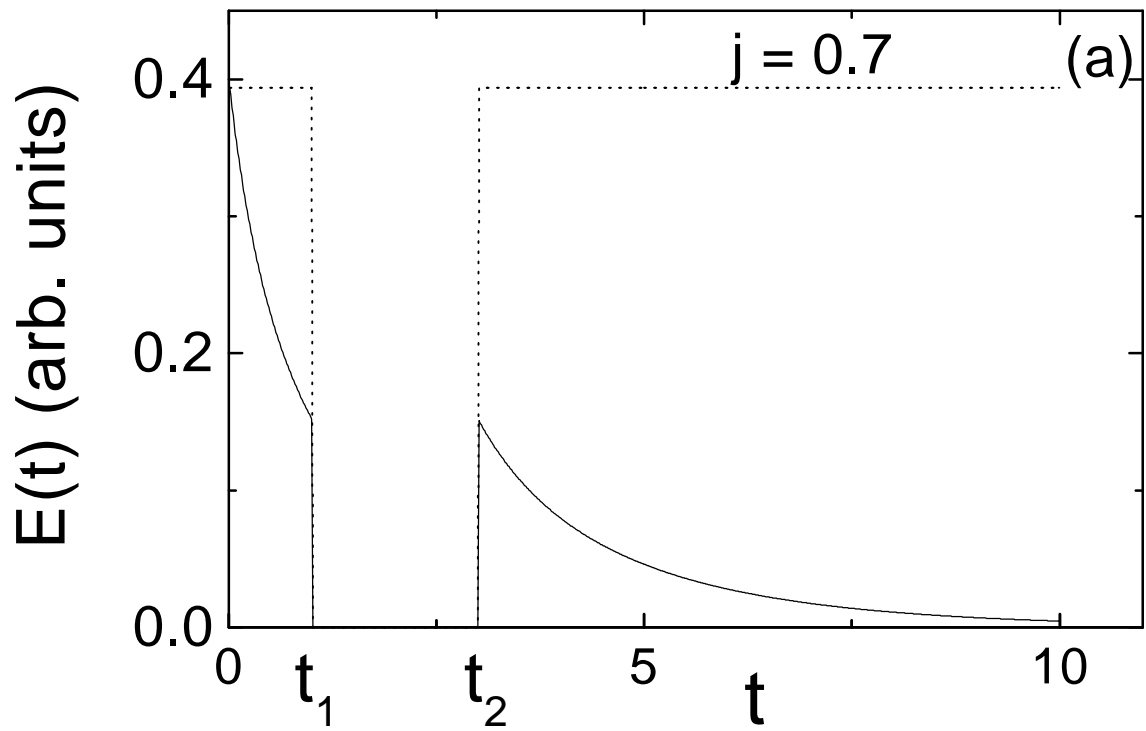
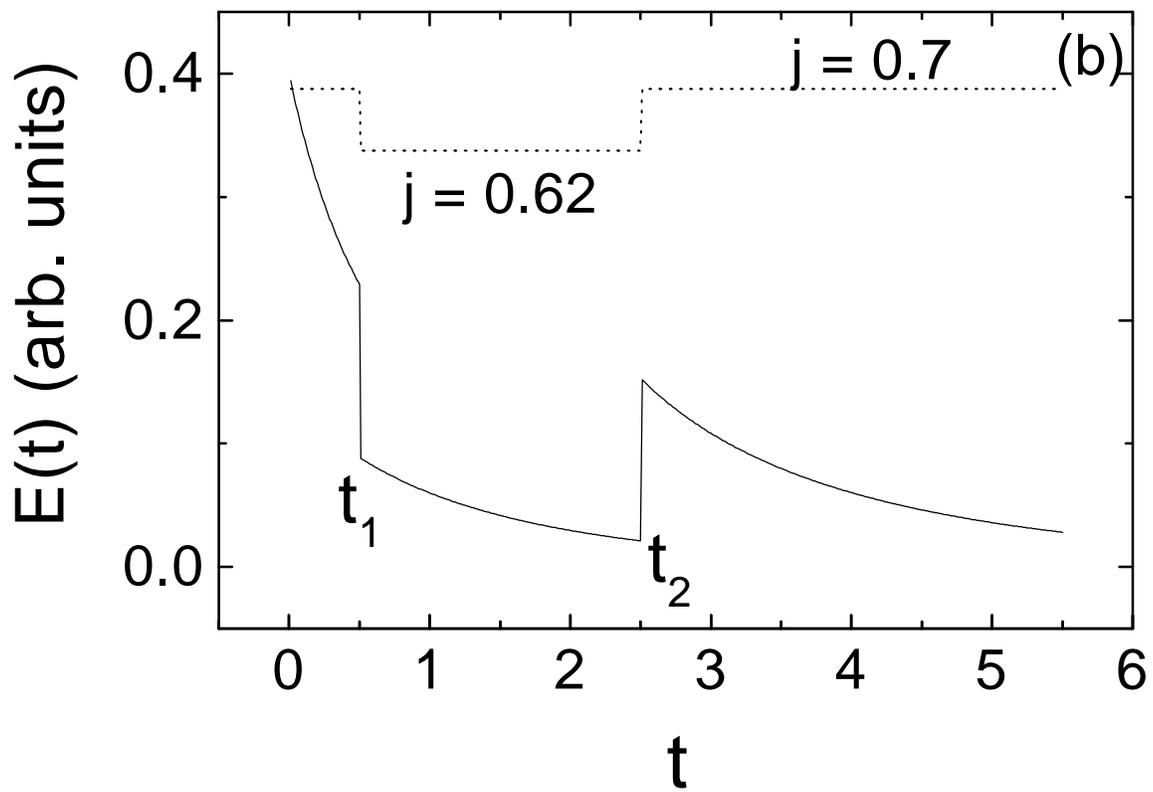
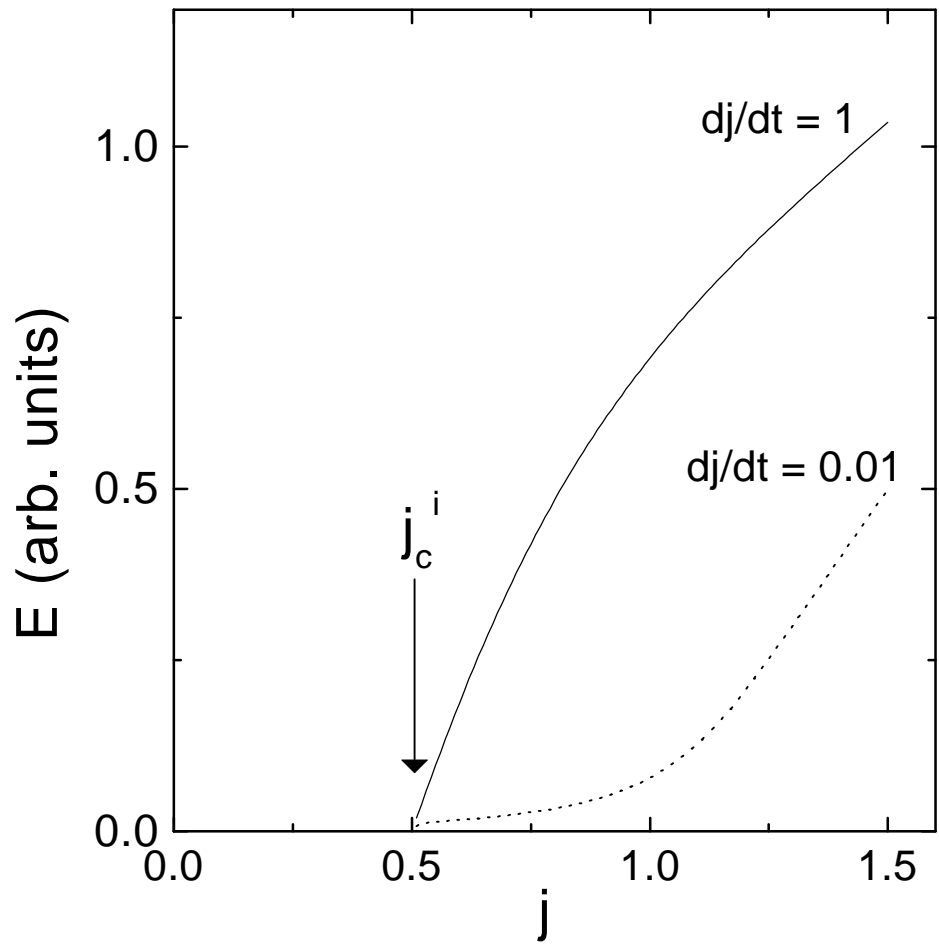
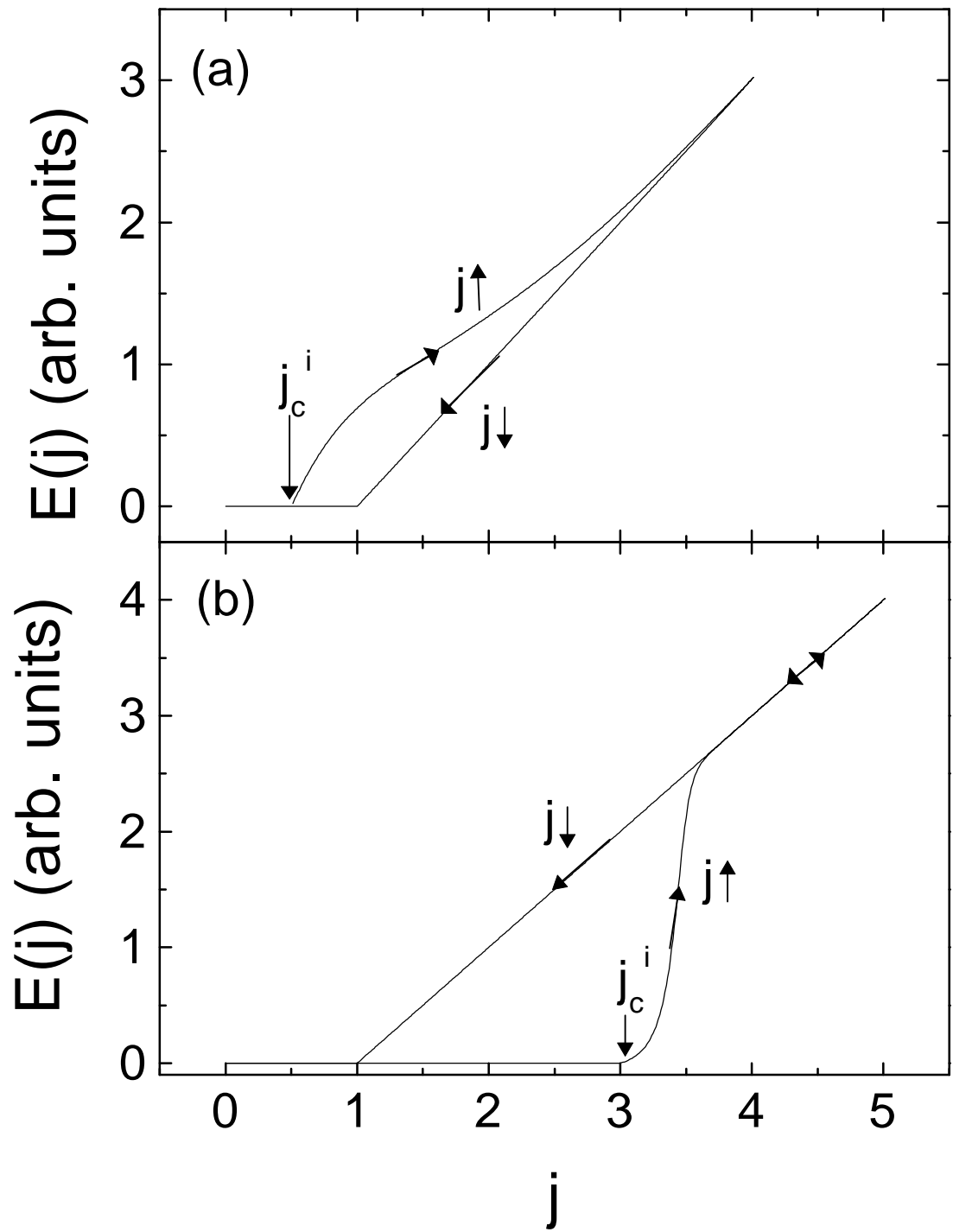
\end{document}